\documentclass[journal]{IEEEtran}

\usepackage{cite}
\usepackage{amsmath,amssymb,amsfonts}
\usepackage{epsfig}
\usepackage{graphics}
\usepackage{textcomp}
\def\BibTeX{{\rm B\kern-.05em{\sc i\kern-.025em b}\kern-.08em
    T\kern-.1667em\lower.7ex\hbox{E}\kern-.125emX}}

\usepackage[table,xcdraw]{xcolor}
\usepackage{algorithm, algpseudocode}

\usepackage{url}

\usepackage{booktabs}

\usepackage{amsthm}

\allowdisplaybreaks

\theoremstyle{plain}
\newtheorem{thm}{\protect\theoremname}
\theoremstyle{plain}
\newtheorem{lem}[thm]{\protect\lemmaname}

\makeatother

\usepackage{babel}
\providecommand{\lemmaname}{Lemma}
\providecommand{\theoremname}{Theorem}

\begin{document}

\title{GOSPA-Driven Non-Myopic Multi-Sensor Management with Multi-Bernoulli Filtering}

\author{George Jones and \'Angel F. Garc\'ia-Fern\'andez\\
\thanks{G. Jones is with Dept. of Autonomy, Frost Unmanned AB, Gothenburg 444 32, Sweden. A. F. Garc\'ia-Fern\'andez is with the IPTC, ETSI de Telecomunicaci\'on, Universidad Polit\'ecnica de Madrid (emails: george.jones@frostunmanned.com, angel.garcia.fernandez@upm.es). 
This work was supported by the EPSRC Centre for Doctoral Training in Distributed Algorithms EP/S023445/1 at the University of Liverpool and Roke Manor Research Limited. This work was partially supported by the Spanish Ministry of Science, Innovation and Universities under the project PID2024-158149OB-C21.}}

\maketitle

\begin{abstract}
In this paper, we propose a non-myopic sensor management algorithm for multi-target tracking, with multiple sensors operating in the same surveillance area. The algorithm is based on multi-Bernoulli filtering and selects the actions that solve a non-myopic minimisation problem, where the cost function is the mean square generalised optimal sub-pattern assignment (GOSPA) error, over a future time window. For tractability, the sensor management algorithm actually uses an upper bound of the GOSPA error and is implemented via Monte Carlo Tree Search (MCTS). The sensors have the ability to jointly optimise and select their actions with the considerations of all other sensors in the surveillance area. The benefits of the proposed algorithm are analysed via simulations.
\end{abstract}


\section{Introduction}
\IEEEPARstart{S}{ensor} management can be defined as the dynamic re-tasking of agile sensors to achieve an operational objective \cite{Hero_book08}. Sensors can be agile in a multitude of ways, from physically repositioning, changing direction or selecting a sensing mode.

Myopic sensor management, sometimes called greedy sensor management, optimises the sensor resources for the immediate benefit of the system, not considering the long term effects of the actions being selected now. Non-myopic sensor management operates on the policy of considering these long-term effects of the actions selected now. Whilst it has an increased computational demand, non-myopic planning often produces more desirable results \cite{Hernandez2023}, \cite{Kreucher2004}.

Sensor management is applicable to many domains, from space surveillance in \cite{Oakes2022}, environmental field estimation in \cite{roy2016} and the general control of multiple sensors in \cite{Wang2018}. This paper deals with multi-sensor management for multiple-target tracking (MTT). In this context the objective is to track the targets as accuratley as possible whilst minimising the number of missed and false targets. One approach to tackle this sensor management problem is via reinforcement learning where a policy can be learnt off-line via simulation \cite{Kreucher06,Huang24,Oakes2022}. This approach has the benefit of not requiring a complete model of target dynamics and measurements at the expense of lower interpretability and a more complex training process. 

In this work, we pose the MTT multi-sensor management problem using a partially observed Markov decision process (POMDP) \cite{Krishnamurthy_book16}, in which the hidden state is a set of targets. Solving the POMDP requires a Bayesian solution to the MTT problem, which is based on target birth/death models, target dynamic model and measurement model. In addition, a POMDP requires a cost function to minimise, or a reward function to maximise.

Sensor management for MTT can be approached in a variety of ways, including reward functions based on the use of information theory. These methods derive information theoretic approaches, utilising divergence measures such as the R\'enyi divergence \cite{Kreucher07, Ristic11b, Kreucher2005}, the Kullback-Leibler divergence (KLD) \cite{LeGrand2023, Saucan2021, Aoki11} and the Cauchy-Schwarz divergence \cite{Beard17b, Gostar2020}. Their objective can broadly be understood to be to take the action that provides the largest gain in information about the variables of interest. Whilst these approaches often produce desirable behaviours, they are not explicitly transparent w.r.t. what the resulting policy is optimising for in terms of the real world.

Another popular cost function is the posterior Cram\'{e}r-Rao lower bound (PCRLB) \cite{Tichavsky1998, Hernandez2006, Tharmarasa2007, Bell15, Tang2019}. The PCRLB is a bound on the mean square error and is used instead of the mean square error because of its good performance and computational efficiency. However, to use the PCRLB as a cost function, external criteria must be added to be able to perform tracking of an time varying and unknown number of targets. In \cite{BostroemRost2021} \cite{BostroemRost2022}, the cost function is the weighted sum of two terms: the tracking cost (the sum of the covariance matrix of each track) and the search cost (the expected number of targets that remain undetected). 

In this paper, we utilise the generalised optimal sub pattern assignment (GOSPA) metric \cite{Rahmathullah2017}, or specifically an upper bound on the mean square GOSPA (MSGOSPA) error as a cost function to inform sensor actions and drive the sensor management \cite{Jones2024TAES}, \cite{Hernandez2023}. By using the GOSPA metric, we are selecting actions which specifically minimise the localisation, missed target and false target errors - the three quantities of interest in a multi-target estimation scenario. Therefore, the GOSPA cost function is a natural choice for planning sensor actions in search and track operations. A comparison between the GOSPA metric and other multi-target metrics for sensor management is given in \cite{GarciaFernandez2021}. An important benefit of GOSPA w.r.t. the optimal sub-pattern assignment (OSPA) metric \cite{Schuhmacher2008} and the unnormalised OSPA metric, which is proportional to the Cardinalized Optimal Linear Assignment metric \cite{Barrios2015}, is that the cost is additive for independent, far-away targets. This property is key to avoiding the spooky effect at a distance in optimal estimation and simplifies the planning stage \cite{GarciaFernandez2021}.

The predecessors to this paper cover the utilisation of the GOSPA metric to manage a single sensor using a (Gaussian) Bernoulli filter for myopic and non-myopic sensor management \cite{Jones2024TAES} and then a multi-Bernoulli filter for myopic sensor management \cite{Jones2024MFI}. This paper has two main contributions: 

\begin{itemize}
    \item A non-myopic, multi-sensor management algorithm for MTT driven by the GOSPA metric based on Gaussian multi-Bernoulli filtering extending the work in \cite{Jones2024MFI} to handle the non-myopic case and multiple sensors.
    \item An efficient implementation of the algorithm using Monte Carlo Tree Search (MCTS) \cite{Browne2012}. For the efficient MCTS implementation, we use hypothesis reduction by merging the detection and misdetection hypotheses for each target such that the edges of the tree only consider sensor actions, and not measurements, as done in \cite{Salvagnini2015}.
\end{itemize}
Simulation results show the benefits of the proposed sensor management algorithm.

The remainder of this paper is organised as follows. In Section \ref{sec: problem formulation}, we provide the background and the objective of this paper. In Section \ref{sec: sensor management}, we introduce the multi-sensor management algorithm for multi-Bernoulli filtering. In Section \ref{sec: MCTS}, we explain the MCTS implementation. In Section \ref{sec: simulations}, we analyse the simulation results of the sensor management algorithm. In Section \ref{sec: conclusion}, we provide the concluding remarks.

\section{Background \& Objective}\label{sec: problem formulation}
This section presents the background and the objective of this paper. In Section \ref{sec: Dynamic and measurement model}, we present the dynamic and measurement model. In Section \ref{sec: multi-Bernoulli filtering}, we present the multi-Bernoulli filter, including its prediction and update step. In Section \ref{sec: GOSPA metric}, we 
present the GOSPA metric and, in Section \ref{sec: non myopic planning using time discounted predicted msgospa error}, we formulate the problem of non-myopic planning, using the time discounted predicted mean square GOSPA error. The main notation in the paper is provided in Table \ref{tab:Notation}.

\begin{table}

\caption{Main notation}\label{tab:Notation}
\rule[0.5ex]{1\columnwidth}{1pt}
\begin{itemize}
\item Filter
\begin{itemize}
\item $f_{k|k}^{s}(\cdot)$: MBM density of the set of targets at time step
$k$ after the $s$-th sensor measurement update at time step $k$. 
\item $f_{k|k-1}^{m,s}\left(\cdot\right)$: predicted measurement density
for the $s$-th sensor.
\item $f_{k|k}^{i}\left(\cdot;a_{k}\right)$: $i$-th Bernoulli density.
\item $r_{k'|k}^{i}$, $\overline{x}_{k'|k}^{i}$, $P_{k'|k}^{i}$: existence
probability, mean and covariance at time step $k'\in\{k,k+1\}$ given
the measurements up to time step $k$ for the $i$-th Bernoulli.
\end{itemize}
\item Non-myopic planning
\begin{itemize}
\item $a_{k:k'}$: sequence of actions from time step $k$ to $k'$.
\item $h_{k:k'}^{i}$: sequence of detections/misdetections for all sensors
from time step $k$ to $k'$ for the $i$-th Bernoulli (Section III.F).
\item $r_{k'|k',a_{k:k'}}^{i,h_{k:k'}^{i}}$, $\overline{x}_{k'|k',a_{k:k'}}^{i,h_{k:k'}^{i}}$,
$P_{k'|k',a_{k:k'}}^{i,h_{k:k'}^{i}}$: existence probability, mean
and covariance at time step $k'$ given $h_{k:k'}^{i}$ and $a_{k:k'}$
for the $i$-th Bernoulli.
\item $p(h_{k'}^{i}|a_{k:k'-1},h_{k:k'-1}^{i})$: probability of the detection/misdetections
$h_{k'}^{i}$ (for all sensors) at time step $k'$ given previous
 $a_{k:k'-1}$ and $h_{k:k'-1}^{i}$.
\item $\mathrm{C}^{i}\left(a_{k:k'}h_{k:k'}^{i}\right)$: cost for the $i$-th
Bernoulli at time step $k'$ given $a_{k:k'}$ and $h_{k:k'}^{i}$. 
\end{itemize}
\item Non-myopic planning with tree reduction
\begin{itemize}
\item $r_{k'|k',a_{k:k'}}^{i}$, $\overline{x}_{k'|k',a_{k:k'}}^{i}$, $P_{k'|k',a_{k:k'}}^{i}$:
merged existence probability, mean and covariance at time step $k'$
given $a_{k:k'}$ for the $i$-th Bernoulli.
\item $r_{k'|k',a_{k:k'}}^{i,h_{k'}^{i}}$, $\overline{x}_{k'|k',a_{k:k'}}^{i,h_{k'}^{i}}$,
$P_{k'|k',a_{k:k'}}^{i,h_{k'}^{i}}$: pre-merged existence probability,
mean and covariance at time step $k'$ given $a_{k:k'}$ and
$h_{k'}^{i}$ for the $i$-th Bernoulli.
\item $p(h_{k'}^{i}|a_{k:k'-1})$: probability of the detections/misdetections
$h_{k'}^{i}$ at time step $k'$ given previous actions $a_{k:k'-1}$.
\item $\mathrm{C}^{i}\left(a_{k:k'},h_{k'}^{i}\right)$: cost for the $i$-th
Bernoulli at time step $k'$ given $a_{k:k'}$ and $h_{k'}^{i}$.
\item $\mathrm{C}_{k'}(a_{k:k'})$: cost at time step $k'$. 
\end{itemize}
\end{itemize}
\rule[0.5ex]{1\columnwidth}{1pt}
\end{table}

\subsection{Dynamic \& Measurement Model}\label{sec: Dynamic and measurement model}
A single target state is denoted as $x \in \mathbb{R}^{n_x}$. The set of targets at time step $k$ is $X_k$ such that  $X_k \in \mathcal{F}(\mathbb{R}^{n_x})$ where $\mathcal{F}(\mathbb{R}^{n_x})$ is the set of all finite subsets of $\mathbb{R}^{n_x}$ \cite{Mahler_book14}. In this paper, we consider that there are $S$ sensors. At time step $k$, there are a discrete number of actions $a_k = (a_k^1, ... ,a_k^S) \in \mathbb{A}_k$ where $a_k^s$ represents the action associated with the $s$-th sensor. For sensor $s$, a target with state $x$ in the surveillance area can either be detected with probability of detection $p_{a_{k}}^{D,s}(x)$ and one measurement is generated  with density $l^s(z|x)=\mathcal{N}\left(z;H^s_{a_{k}}x+b^s_{a_{k}},R^s_{a_{k}}\right)$, or missed (not detected) with probability $1 - p_{a_{k}}^{D,s}(x)$. At time step $k$, the set of measurements from sensor $s$, which is denoted by $Z_k^s$, can either be empty $Z_k^s=\emptyset$ or contain multiple measurements $|Z_k^s| \geq 1$, where measurements are the union of the target-generated detections and Poisson point process (PPP) clutter, defined by intensity $\lambda^C_{a_k}(\cdot)$. The sequence of sets of measurements from each sensor received at time step $k$ is denoted as $Z_k = (Z_k^1, ... , Z_k^S)$.

The dynamic model of the targets is as follows. The probability of a target with state $x$ surviving and transitioning to the next time step is $p^S(x)$ with the single-target transition density being $g(\cdot|x)$, or the target can die with probability $1-p^S(x)$. The multi-target state $X_{k+1}$ is given by the union of the new and surviving targets.

We consider a multi-Bernoulli (MB) birth model which contains $n^b_k$ Bernoulli components. The $i$-th Bernoulli component has an associated probability of birth $r^{b,i}_k$ and single-target density $p^{b,i}_k(\cdot)$. The density is given by \cite{Mahler_book14,GarciaFernandez2018}
\begin{equation}\label{eqn: bernoullis birth}
    f^b_k(X_k) = \sum_{X^1 \uplus...\uplus X^{n^b_k}= X_k} \prod_{i=1}^{n^b_k}f_k^{b,i}(X^i)
\end{equation}
where the $i$-th Bernoulli component density is
\begin{equation}
    f_k^{b,i}(X_k) = 
    \begin{cases}
        1 - r_k^{b,i} & X_k = \emptyset \\
        r_k^{b,i}p_k^{b,i}(x) & X_k = \{x\} \\
        0 & \mathrm{otherwise.}
    \end{cases}
\end{equation}

The summation in \eqref{eqn: bernoullis birth} is taken over all mutually disjoint and (potentially) empty sets $X^1, ... ,X^{n^b_k}$ whose union is $X_k$.

\subsection{Multi-Bernoulli Filtering}\label{sec: multi-Bernoulli filtering}
For the previous dynamics and measurement models, the posterior density is a multi-Bernoulli mixture (MBM), which can be computed in closed-form via the MBM filtering recursion \cite{GarciaFernandez2018}, \cite{GarciaFernandez2019} performing $S$ sequential updates (one per sensor) at each update step.

In this paper for simplicity, we consider a multi-Bernoulli filter. In MB filtering, both the predicted and posterior densities at time step $k' \in \{k,k+1\}$, given the measurements up to time step $k$, are approximated as an MB of the form
\begin{equation}\label{eqn: MB form}
    f_{k'|k}(X_{k'}) = \sum_{X^1 \uplus...\uplus X^{n_{k'|k}}= X_{k'}} \prod_{i=1}^{n_{k'|k}}f_{k'|k}^{i}(X^i)
\end{equation}
where $n_{k'|k}$ is the number of Bernoulli components.

For the $i$-th Bernoulli $f^i_{k'|k}(\cdot)$, the two quantities that must be propagated over time via the prediction and update steps are the spatial density $p_{k'|k}^{i}(\cdot)$ and the probability of existence $r_{k'|k}^{i}$. 
%
%
\subsubsection{MB Filter Prediction Step}
The predicted density has $n_{k+1|k} = n_{k|k} + n^b_{k+1}$ Bernoulli components, including the previous and new Bernoulli components. The prediction equations for Bernoulli components relating to surviving targets are \cite{GarciaFernandez2020}
\begin{equation}
    r^i_{k+1|k} = r^i_{k|k} \langle p^S, p^i_{k|k}\rangle
\end{equation}
\begin{equation}\label{eqn: bernoullis target}
    p^i_{k+1|k}(x) = \frac{\int g(x|y)p^S(y)p^i_{k|k}(y)dy}{\langle p^i_{k|k},p^S\rangle}
\end{equation}
where the notation $\langle h,g \rangle $ denotes the inner product between two functions $h(\cdot)$ and $g(\cdot)$ \cite{Mahler_book14}.

For Bernoulli components representing new born targets, $i \in \{n_{k|k}+1, ... , n_{k|k}+n^b_{k+1}\}$, the parameters are those of the birth model \eqref{eqn: bernoullis birth}, which results in
\begin{equation}\label{eqn: birth prob}
    r^i_{k+1|k} = r^{b,i-n_{k|k}}_{k+1}
\end{equation}
\begin{equation}\label{eqn: birth density}
    p^i_{k+1|k}(x) = p^{b,i-n_{k|k}}_{k+1}(x).
\end{equation}
\subsubsection{MB Filter Update Step}
To obtain a multi-sensor MB filter we can apply $S$ MB sequential updates, one for each sensor. After each update, the density is of MBM form \cite{GarciaFernandez2019}, 
\begin{equation}\label{eqn: MBM form}
    f^s_{k|k}(X_k)=\sum_{e \in \mathbb{E}_{k|k}} w^{s,e}_{k|k} \sum_{\uplus_{l=1}^{n_{k|k}}X^l = X_{k}} \prod_{i=1}^{n_{k|k}}f^{i,s,e^i}_{k|k}(X^i)
\end{equation}
where $\mathbb{E}_{k|k}$ is the set that represents all global data association hypotheses and $e = (e^1, ... , e^{n_{k|k}})$, $e^i$ is the local hypothesis for Bernoulli component $i$. It should be mentioned that the standard notation for global hypotheses in Poisson multi-Bernoulli mixture filtering, see \cite{Williams2015} \cite{GarciaFernandez2019}  for details, is $a$ and has not been used here as $a$ is used to denote sensor actions. We proceed to explain the update of the local hypothesis.

Given the measurement set of the $s$-th sensor $Z_k^s = \{z_k^{s,1}, ... ,z_k^{s, m_k^s}\}$, the $s$-th update equations for a misdetection hypothesis for the $i$-th Bernoulli component are
\begin{equation}\label{eqn: updated weight}
    w_{k|k}^{i,s,1} = 1 - r_{k|k}^{i,s-1} \langle p_{k|k}^{i,s-1},p^{D,s}_{a_k} \rangle
\end{equation}
\begin{equation}
    r^{i,s,1}_{k|k} = \frac{r^{i,s-1}_{k|k}\langle p^{i,s-1}_{k|k}, 1-p^{D,s}_{a_k} \rangle}{1 - r^{i,s-1}_{k|k} \langle p^{i,s-1}_{k|k},p^{D,s}_{a_k}\rangle}
\end{equation}
\begin{equation} \label{eqn: updated single target density}
    p_{k|k}^{i,s,1}(x) = \frac{(1-p^{D,s}_{a_k}(x))p_{k|k}^{i,s-1}(x)}{\langle p_{k|k}^{i,s-1}, 1-p^{D,s}_{a_k} \rangle}
\end{equation}
where the superindex $(i,s,j)$ denotes the $i$-th Bernoulli component, the $s$-th sensor and the $j$-th local hypothesis, with $j=1$ corresponding to a misdetection.

The equations for the update of the detection hypothesis for Bernoulli component $i$ and measurement $z_k^{s,j}$ are given by
\begin{equation}\label{eqn: detection hypothesis w}
    w_{k|k}^{i,s,1+j} = \frac{r_{k|k}^{i,s-1} \langle p_{k|k}^{i,s-1}, l^s(z_k^{s,j}|\cdot)p^{D,s}_{a_k} \rangle}{\lambda^C_{a_k}(z^{s,j}_k)}
\end{equation}
\begin{equation}
    r_{k|k}^{i,s,1+j} = 1
\end{equation}
\begin{equation}
    p_{k|k}^{i,s,1+j}(x) = \frac{l^s(z_k^{s,j}|x)p^{D,s}_{a_k}(x)p_{k|k}^{i,s-1}(x)}{\langle p_{k|k}^{i,s-1}, l^s(z_k^{s,j}|\cdot)p^{D,s}_{a_k} \rangle}.
\end{equation}

These update equations give rise to the updated density after the $s$-th sensor update, which is of MBM form \eqref{eqn: MBM form}. 

After each update, the MBM is projected to an MB, via Kullback-Leibler divergence (KLD) minimisation with auxiliary variables \cite{GarciaFernandez2020, Kim2024}. After the $s$-th update, this MB density is characterised by
\begin{equation}\label{eqn: MBM KLD prob existence}
    r^{i,s}_{k|k} = \sum_{e^i=1}^{h^i}{w}^{i,s,e^i}_{k|k}r^{i,s,e^i}_{k|k},
\end{equation}
\begin{equation}
p^{i,s}_{k|k}(x)=\frac{\sum_{e^i=1}^{h^i}{w}^{i,s,e^i}_{k|k}r^{i,s,e^i}_{k|k}p^{i,s,e^i}_{k|k}(x)}{\sum_{e^i=1}^{h^i}{w}^{i,s,e^i}_{k|k}r^{i,s,e^i}_{k|k}},
\end{equation}
\begin{equation} \label{eqn: MBM KLD weight}
    {w}^{i,e^i}_{k|k} = \sum_{b \in \mathbb{E}_{k|k}:b^i=e^i} w^b_{k|k}
\end{equation}
where $h^i=1+m_{k}^s$ is the number of local hypotheses of the $i$-th Bernoulli \cite{GarciaFernandez2019}. We also consider that $r^{i,0}_{k|k} = r^{i}_{k|k-1}$ and $p^{i,0}_{k|k}(\cdot) = p^{i}_{k|k-1}(\cdot)$.

There are many methods that can approximately perform this MB projection in a computationally efficient manner, for instance Murty's algorithm \cite{Murty1968}, loopy belief propagation \cite{Williams2014}, \cite{Meyer2018}, \cite{Kim2024} efficient hypothesis management \cite{Horridge2006} and sampling based approaches \cite{Morelande2009}.


\subsection{The GOSPA metric}\label{sec: GOSPA metric}
The GOSPA metric \cite{Rahmathullah2017} is a metric for sets of targets that penalises the localisation error, the missed detection error and the false detection error. Let $X=\{x_1,...,x_{|X|}\}$ be the ground truth set of targets and $Y=\{y_1,...,y_{|Y|}\}$ be the estimated set. Let $\gamma$ be an assignment set from $X$ to $Y$. The assignment set $\gamma$ has these properties  $\gamma \subseteq \{ 1,...,|X|\} \times \{1,...,|Y|\}, (i,j), (i,j') \in \gamma \Longrightarrow j=j' \quad \text{and} \quad (i,j),(i',j) \in \gamma \Longrightarrow i = i'$. These properties imply that each target in $X$  and $Y$ can be assigned at most once. Then, given a single-target metric $d(\cdot, \cdot)$ parameter $p > 0$ and $\alpha = 2$, the GOSPA metric between set $X$ and set $Y$ is
\begin{multline}\label{eqn: gospa metric}
    d (X,Y) = \\
     \underset{\gamma\in\Gamma}{\text{min}}\left( \sum\limits_{(i,j) \in \gamma} d^p(x_i,y_j) + \frac{c^p}{2}(|X| + |Y| - 2|\gamma|)\right)^{1/p}.
\end{multline}

In the remainder of the paper, we use $p=2$.

\subsection{Non-myopic Planning with Time-discounted MSGOSPA Error} \label{sec: non myopic planning using time discounted predicted msgospa error}
In this paper, the objective is to plan the sensor actions by minimising the MSGOSPA error in a window of $T$ time steps from time step $k$ to time step $K=k+T-1$. Sensor management consists of determining a policy $\mu_{k'}(\cdot)$ at time step $k'$ that maps the available information $\mathcal{I}_{k'}$ at time step $k'$ to an action at time step $k'$. The available information at time step $k'$ is 
\begin{equation}
\mathcal{I}_{k'} = 
    \begin{cases}
        (f_{k|k-1}(\cdot)) & k'= k \\
        (f_{k'|k'-1}(\cdot), a_{k:k'-1}, Z_{k:k'-1}) & k'\geq k
    \end{cases}
\end{equation}
such that $a_{k'} = \mu_{k'}(\mathcal{I}_{k'})$.

The sequence of policies (which are the deterministic functions that map information to actions) up to the planning horizon is denoted by $\mu_{k:K}(\cdot) = (\mu_{k}(\cdot), \ldots, \mu_{K}(\cdot))$. Including a decay factor (discount rate) of $\lambda \in (0,1]$, the policy is chosen to minimise \cite[Eq. (7.6)]{Krishnamurthy_book16} 
\begin{multline}\label{eqn: J not nested}
    J_{\mu_{k:K}(\cdot)} = \\
    E_{\mu_{k:K}(\cdot)}\left[ \sum^K_{k'=k} \lambda^{k'-k}d^2\left(X_{k'}, \hat{X}_{k'}(a_{k:k'}, Z_{k:k'})\right)\right]
\end{multline}
where the expectation is taken with respect to the joint probability density of $(X_k, \ldots, X_K, Z_k,\ldots,Z_K)$ under the policy $\mu_{k:K}(\cdot)$, and $\hat{X}_{k'}(a_{k:k'},Z_{k:k'})$ is the optimal MSGOSPA estimator at time step $k'$.

The optimisation over the policy in \eqref{eqn: J not nested} can be written using a value function that is propagated backwards using a Bellman-type equation \cite[Chapter 7]{Krishnamurthy_book16}, see (12)-(13) in \cite{Jones2024TAES}. 
%
%
%
The objective of this paper is to develop a computationally efficient approximation to solve \eqref{eqn: J not nested} based on MB filtering.

\section{Multi-sensor Management}\label{sec: sensor management}
This section describes an algorithm for managing multiple sensors using multi-Bernoulli filtering, using the Gaussian distribution and the GOSPA metric.

As described in Section \ref{sec: non myopic planning using time discounted predicted msgospa error}, the objective is to perform planning by minimising a time-discounted MSGOSPA error, see \eqref{eqn: J not nested}. Since calculating the MSGOSPA error is intractable, we instead use an upper bound that we proceed to calculate. To do so, in Section \ref{sec: assumptions of the sensor management algorthm}, we state the assumptions used in the sensor management algorithm. We compute the density of the measurements in Section \ref{sec: measurement density}. We then derive an upper bound to the overall MSGOSPA error for a given MB posterior in Section \ref{sec: MSGOSPA error for given MB posterior}. This result is then used to derive an upper bound on the MSGOSPA error in the myopic case in Section \ref{sec: MSGOSPA error computation}. In Section \ref{sec: multi sensor update}, we explain the multi-sensor update that is required for sensor management and finally, in Section \ref{sec: bellman eqn}, we provide the resulting non-myopic planning algorithm.

\subsection{Assumptions of the Sensor Management Algorithm}\label{sec: assumptions of the sensor management algorthm}
This section presents the assumptions of the sensor management algorithm. In the MB filter, we consider a Gaussian implementation in which each single-target density is Gaussian with density
\begin{equation}
    p^{i}_{k'|k}(x_{k'}) = \mathcal{N}(x_{k'};\overline{x}_{k'|k}, P_{k'|k})
\end{equation}
where $\overline{x}_{k'|k}$ and $P_{k'|k}$ denote the mean and covariance matrix. The sensor management algorithm also makes the following assumptions:

\begin{enumerate}
    \item We either detect zero measurements $Z^s_k = \emptyset$ or one measurement $Z^s_k = \{z_k\}$ at the predicted measurement for each Bernoulli component, without clutter measurements. \label{assumption 2}
    \item The probability of detection is approximated as a constant given by its predicted value for each Bernoulli component. \label{assumption 4}
    \item We use a computationally efficient upper bound for the resulting MSGOSPA error (see Section \ref{sec: MSGOSPA error for given MB posterior}). \label{assumption 5}
    \item The sensors are operating in a centralised manner, meaning they are operating with the same prediction information at each time step. \label{assumption 6}
    \item All Bernoulli components are far away from each other.\label{assumption 7}
    \item The measurements of the $S$ sensors at a given time step are independent. \label{assumption 8}
\end{enumerate}

It should be noted that these assumptions are only made by the sensor management algorithm but not by the MB filter that is run once we have taken the actions and received the measurements. It is also relevant to notice that in \cite{BostroemRost2021}, the sensor management algorithm also assumes that there is no clutter, but always considers a detection for each target. Considering misdetections as in Assumption \ref{assumption 2}) is more general and can become important for low probabilities of detection. In addition, while Assumptions 1)-5) can be strong approximations in some situations, they enable the development of a closed-form upper bound of the MSGOSPA error for sensor management with MB filters that is computationally efficient to implement.

\subsection{Measurement Density}\label{sec: measurement density}
The predicted measurement density for an MB predicted density is Poisson multi-Bernoulli \cite{Williams2015a}. This is due to the fact that the clutter measurements follow a PPP and each potential target generates a potential measurement that follows a Bernoulli distribution. The expected probability of detection for the $s$-th sensor and $i$-th Bernoulli component with mean $\overline{x}^{i,s}_{k|k-1}$ and covariance $P^{i,s}_{k|k-1}$ is given by 
\begin{align}\label{eqn: expected pd}
    \overline{p}_{a_{k}}^{D,s,i} 
    & = \int {p}_{a_{k}}^{D,s}\left(x\right)\mathcal{N}\left(x;\overline{x}_{k|k-1}^i,P_{k|k-1}^i\right)dx.
\end{align}
For an MB predicted density, the density of the predicted sequence of measurements at time step $k$ under Assumption \ref{assumption 8}) is
\begin{align} \label{eqn: predicted density of measurement sequence}
    f_{k|k-1}^{m}\left(Z_{k}\right) & \approx \prod_{s=1}^{S}f_{k|k-1}^{m,s}\left(Z_{k}^{s}\right)
\end{align}
where
\begin{multline}
    f_{k|k-1}^{m,s}\left(Z_{k}^{s}\right) = \\
    \sum_{\uplus_{l=0}^{n_{k|k-1}}Z^{s,l}=Z_{k}^{s}}f_{k|k-1}^{m,p,s}\left(Z^{s,0}\right)
    \prod_{i=1}^{n_{k|k-1}}f_{k|k-1}^{m,i,s}\left(Z^{s,i}\right)\label{eq:predicted_measurement}
\end{multline}
is a Poisson multi-Bernoulli density as $f_{k|k-1}^{m,p,s}\left(Z^{s,0}\right)$ is a PPP density with intensity $\lambda^C_{a_k}(\cdot)$ and under Assumption \ref{assumption 4}) and using \eqref{eqn: expected pd}, $f_{k|k-1}^{m,i,s}\left(Z^{s,i}\right)$ is a Bernoulli density given by 
\begin{multline}\label{eqn: predicted density of the measurement}
    f_{k|k-1}^{m,i,s}({Z}_k^{s,i};a_k) = \\
    \begin{cases}
      \overline{p}^{D,s,i}_{a_k} r_{k|k-1}^{i} \mathcal{N}(z_k^s;, \hat{z}_{a_k}^{s,i},S_{a_k}^{s,i}) & {Z}_k^{s,i} = \{z_k^s\}\\
      1 - r_{k|k-1}^{i} \overline{p}^{D,s,i}_{a_k} & {Z}_k^{s,i} = \emptyset 
    \end{cases}
\end{multline}
where $\hat{z}_{a_k}^{s,i}$ and $S_{a_k}^{s,i}$ are the Kalman filter predicted measurement and its covariance matrix (for the $s$-th sensor and the $i$-th Bernoulli) \cite{Sarkka_book23}.

\subsection{MSGOSPA Error for a Given MB Posterior}\label{sec: MSGOSPA error for given MB posterior}
Applying Lemma 2 in \cite{Williams2015}, the MSGOSPA error at time step $k$ for a given MB posterior $f_{k|k}(\cdot)$ is
\begin{align} \label{eqn: MSGOSPA for given MB posterior}
     & \int d^{2}\left(X_{k},\hat{X}_{k}\left(a_{k},Z_{k}\right)\right)f_{k|k}\left(X_{k}|Z_{k};a_{k}\right)\delta X_{k} \nonumber \\
     & =\int d^{2}\left(\cup_{i=1}^{n_{k|k}}X^{i},\cup_{i=1}^{n_{k|k}}\hat{X}^{i}\left(a_{k},Z_{k}^{i}\right)\right) \nonumber\\
     &\quad\quad\quad\quad\quad\quad\quad\quad \prod_{i=1}^{n_{k|k}}f_{k|k}^{i}\left(X^{i};a_{k}\right)\delta X^{1:n_{k|k}}
\end{align}
where $\hat{X}^{i}$ is the set estimate (which may be empty or contain a single state) of the $i$-th Bernoulli and $X^{1:n_{k|k}} = (X^1,...,X^{n_{k|k}})$. Under Assumption \ref{assumption 7}) and the property that the GOSPA metric is additive for far away targets \cite{GarciaFernandez2021}, we obtain that the MSGOSPA error is
\begin{multline}\label{eqn:MSGOSPA_additive}
    \int d^{2}\left(X_{k},\hat{X}_{k}\left(a_{k},Z_{k}\right)\right)f_{k|k}\left(X_{k}|Z_{k};a_{k}\right)\delta X_{k} \\
    \simeq\sum_{i=1}^{n_{k|k}}\int d^{2}\left(X^{i},\hat{X}^{i}\left(a_{k},Z_{k}^{i}\right)\right)f_{k|k}^{i}\left(X^{i};a_{k}\right)\delta X^{i}.
\end{multline}
We consider that a target is detected if its probability of existence is higher than the detection threshold $\Gamma_d$. Then under Assumptions \ref{assumption 2}) and \ref{assumption 7}), we apply the upper bound in \cite[Lem. 1]{Jones2024TAES} for Gaussian Bernoulli components such that 
\begin{multline} \label{eqn: cost inequality bernoulli components}
    \int d^{2}\left(X_{k},\hat{X}_{k}\left(a_{k},Z_{k}\right)\right)f_{k|k}\left(X_{k}|Z_{k};a_{k}\right)\delta X_{k} \\ \leq\sum_{i=1}^{n_{k|k}}\mathrm{C}\left(\Gamma_{d},r_{k|k,a_{k}}^{i,|Z_{k}^{i}|},P_{k|k,a_{k}}^{i,|Z_{k}^{i}|}\right)
\end{multline}
and
\begin{multline}\label{eqn: final cost split}
    \mathrm{C}\left(\Gamma_{d},r_{k|k,a_{k}}^{i,|Z_{k}^{i}|},P_{k|k,a_{k}}^{i,|Z_{k}^{i}|}\right) = \\
    \begin{cases}
      \frac{c^2}{2}r_{k|k,a_{k}}^{i,|Z_{k}^{i}|} & r_{k|k,a_{k}}^{i,|Z_{k}^{i}|} \leq \Gamma_d\\
      \frac{c^2}{2}(1 - r_{k|k,a_{k}}^{i,|Z_{k}^{i}|}) \\
      + r_{k|k,a_{k}}^{i,|Z_{k}^{i}|} \min \left(\text{tr}(P_{k|k,a_{k}}^{i,|Z_{k}^{i}|}),c^2\right) &  r_{k|k,a_{k}}^{i,|Z_{k}^{i}|} > \Gamma_d\\
    \end{cases}   
\end{multline}
where we use the notation $|Z_{k}^{i}|=\left(|Z_{k}^{1,i}|,...,|Z_{k}^{S,i}|\right)$ and $r_{k|k,a_{k}}^{i,|Z_{k}^{i}|}$ and $P_{k|k,a_{k}}^{i,|Z_{k}^{i}|}$ are the updated probability of existence and updated covariance of target $i$ with the sequence of detections/misdetections in $|Z^i_k|$.  Note that due to Assumption \ref{assumption 2}), $|Z^i_k|=0$ or $|Z^i_k|=1$.

We can also obtain the optimal detection threshold $\Gamma_d^*$ for the bound \eqref{eqn: final cost split} as \cite{Jones2024TAES}.
\begin{equation}
    \label{eqn: optimal detection threshold}
    \Gamma_d^* = \frac{1}{2 - \min \left(2 \frac{\text{tr}(P_{k|k,a_{k}}^{i,|Z_{k}^{i}|})}{c^2},1\right)}.
\end{equation}

To simplify notation, we define
\begin{equation}\label{eqn: simplified cost notation}
    \mathrm{C}\left(r_{k|k,a_{k}}^{i,|Z_{k}^{i}|},P_{k|k,a_{k}}^{i,|Z_{k}^{i}|}\right)= \mathrm{C}\left(\Gamma_{d}^{*},r_{k|k,a_{k}}^{i,|Z_{k}^{i}|},P_{k|k,a_{k}}^{i,|Z_{k}^{i}|}\right).
\end{equation}

It should be mentioned that Assumption 5) has been used to obtain \eqref{eqn:MSGOSPA_additive}. Under Assumption 5), the optimal assignment problem of GOSPA, see \eqref{eqn: gospa metric}, does not need to be explicitly solved since it is known what estimate corresponds to what target. If targets were in close proximity, this approximation would not be accurate, and one would need to solve the optimal assignment problem. Nevertheless, with targets in close proximity, \eqref{eqn:MSGOSPA_additive} is an upper bound to the MSGOSPA error, which is in any case what is used later in \eqref{eqn: cost inequality bernoulli components}. The reason why it is an upper bound to MSGOSPA is that the assignment is not necessarily the optimal one, so the obtained value is higher.

\subsection{MSGOSPA Error Upper Bound in One Time Step} \label{sec: MSGOSPA error computation}
In this section, we derive an upper bound of the MSGOSPA error in a single time step, which is obtained by setting $K=k$ in \eqref{eqn: J not nested} (myopic planning). The upper bound is provided via the following lemma.

\vspace{5pt}
\hrule\hrule
\begin{lem} 
    Let $r_{k|k,a_{k}}^{i,|Z_{k}^{i}|}$ and $P_{k|k,a_{k}}^{i,|Z_{k}^{i}|}$
    be the updated probability of existence and covariance matrix of the
    $i$-th Bernoulli for the sequence of detections/misdetections $|Z_{k}^{i}|=\left(|Z_{k}^{1,i}|,...,|Z_{k}^{S,i}|\right)$
    where $|Z_{k}^{s,i}|\in\{0,1\}$ with $s\in\{1,...,S\}$. Under Assumptions
    1)-2), 4)-6) , an upper bound on the MSGOSPA error for a given sequence
    $Z_{k}=\left(Z_{k}^{1},...,Z_{k}^{S}\right)$ of measurements at time
    step $k$ is
    
    \begin{align*}
         & \int\left[\int d^{2}\left(X_{k},\hat{X}_{k}\left(a_{k},Z_{k}\right)\right)f_{k|k}\left(X_{k}|Z_{k};a_{k}\right)\delta X_{k}\right] \\
         &\quad\quad\quad\quad\quad\quad\quad\quad\quad\quad\quad\quad\quad\quad \prod_{s=1}^{S}f_{k|k-1}^{m,s}\left(Z_{k}^{s}\right)\delta Z_{k}^{1:S}\\
         & \leq\sum_{i=1}^{n_{k|k}}\sum_{|Z_{k}^{1,i}|=0}^{1}...\sum_{|Z_{k}^{S,i}|=0}^{1}\int\mathrm{C}\left(r_{k|k,a_{k}}^{i,|Z_{k}^{i}|},P_{k|k,a_{k}}^{i,|Z_{k}^{i}|}\right)\\
         & \quad\times\prod_{s=1}^{S}\left[\left(1-|Z_{k}^{s,i}|\right)\left(1-r_{k|k-1}^{i}\overline{p}_{a_{k}}^{D,i}\right)+|Z_{k}^{s,i}|r_{k|k-1}^{i}\overline{p}_{a_{k}}^{D,i}\right].
    \end{align*}
    
    Lemma 1 is proved in Appendix \ref{appendix_A}.
\end{lem}
\vspace{3pt}
\hrule\hrule
\vspace{5pt}
This MSGOSPA upper bound is used by the sensor management algorithm due to its ease of computation.

\subsection{Multi-sensor Update for sensor management}\label{sec: multi sensor update}
This section provides the updated mean and covariance for each Bernoulli in the multi-sensor update required by the sensor management algorithm to calculate the expression given in Lemma 1 . 
The sequential multi-sensor update is computed under Assumptions \ref{assumption 2})-\ref{assumption 4}),  \ref{assumption 6})-\ref{assumption 8}. We encode the sequence of detection/misdetections up to the processing of the $s$-th sensor as $h= (h_1, ... , h_s)$ where $h_s = 0$ if there is a misdetection at sensor $s$ and $h_s = 1$ if there is a detection at sensor $s$. More explicitly, when $h = (0,1)$ this represents a dual-sensor setup, in which sensor 1 is in the misdetection hypothesis and sensor 2 is in the detection hypothesis.

We denote the probability of existence, mean and covariance with sequence of detections and misdetections in $h = (h_1, ... ,h_{s-1})$ taking action $a_k$ for the $i$-th Bernoulli as $r_{k|k,{a_k}}^{i,(h)}, \overline{x}_{k|k,{a_k}}^{i,(h)}, P_{k|k,{a_k}}^{i,(h)}$. Then, the updates with a misdetection and a detection at the $s$-th sensor are the following.
%

For ${Z}_k^{i,s} = \emptyset$, we obtain \cite{GarciaFernandez2019}
\begin{align}
    \overline{x}_{k|k,{a_k}}^{i,(h,0)} &= \overline{x}_{k|k,{a_k}}^{i,h} \label{eqn: x0} \\
    P_{k|k,{a_k}}^{i,(h,0)} &= P_{k|k,{a_k}}^{i,h} \label{eqn: P0} \\
    r_{k|k,{a_k}}^{i,(h,0)} &= \frac{(1-\overline{p}^{D,s,i}_{a_k}) r_{k|k,{a_k}}^{i,h}}{1 - r_{k|k,{a_k}}^{i,h}  + (1 - \overline{p}^{D,s,i}_{a_k})r_{k|k,{a_k}}^{i,h} }. \label{eqn: r0}
\end{align}

For ${Z}_k^{i,s} = \{\hat{z}_{a_k}^{i,s}\}$, there is one measurement at the predicted value. Therefore, the updated mean coincides with the predicted mean. The updated covariance matrix is given by the Kalman filter update, and the updated probability of existence is one \cite{GarciaFernandez2019}
\begin{align}
    \overline{x}_{k|k,{a_k}}^{i,(h,1)} &= \overline{x}_{k|k,{a_k}}^{i,h} \label{eqn: x1} \\
    P_{k|k,{a_k}}^{i,(h,1)} &= P_{k|k,{a_k}}^{i,h} - P_{k|k,{a_k}}^{i,h} ({H^s_{a_k}})^T 
    (S_{a_k}^{i, s,h})^{-1} H^s_{a_k} P_{k|k,{a_k}}^{i,h} \label{eqn: P1} \\
    r_{k|k,a_k}^{i,(h,1)} &= 1. \label{eqn: r1}
\end{align}

It is important to note that the predicted mean is not affected by the value of $h_k^{i}$ or the sequence of actions. Thus, for the mean, we just need the predicted value $\bar{x}_{k'|k}^{i}$.

\subsection{Non-myopic Planning} \label{sec: bellman eqn}
Following Section \cite[Sec. III,E]{Jones2024TAES}, we proceed to write the non-myopic planning problem \eqref{eqn: J not nested} for the MSGOSPA error with the considered upper bound in Lemma 1 and the multi-sensor update. 

Let $h^i_{k:k'}$ denote the sequence of detections/misdetections from time step $k$ to $k'$ for the $i$-th Bernoulli, see Section \ref{sec: multi sensor update}. The probability of existence, mean and covariance matrix of the $i$-th Bernoulli for sequence of actions $a_{k:k'}$ and $h^i_{k:k'}$, computed via \eqref{eqn: x0}-\eqref{eqn: r1}, are $r_{k'|k',a_{k:k'}}^{i,h_{k:k'}^{i}}$, $\overline{x}_{k'|k',a_{k:k'}}^{i,h_{k:k'}^{i}}$ and $P_{k'|k',a_{k:k'}}^{i,h_{k:k'}^{i}}$, respectively.  Then, under the assumptions in Section \ref{sec: assumptions of the sensor management algorthm}, the probability of $h^i_{k'}$ given previous actions and previous sequences of $h^i_{k}$ is
\begin{multline} \label{eqn: observing a measurement density multi-sensor}
    p(h_{k'}^{i}|a_{k:k'},h_{k:k'-1}^{i}) = \\
    \prod_{s=1}^{S}\bigg[\left(1-h_{k'}^{i,s}\right)\left(1-r_{k'|k'-1,a_{k:k'-1}}^{i,h_{k:k'-1}^{i}}\overline{p}_{a_{k:k'}}^{D,s,i}\right)\\
    +h_{k'}^{i,s}r_{k'|k'-1,a_{k:k'-1}}^{i,h_{k:k'-1}^{i}}\overline{p}_{a_{k:k'}}^{D,s,i}\bigg]
\end{multline}

\noindent where $r_{k'|k'-1,a_{k:k'-1}}^{i,h_{k:k'-1}^{i}}$ is the predicted probability of existence at time step $k'$ for the $i$-th Bernoulli and sequence of detection/misdetections in $h_{k:k'-1}^{i}$ and actions $a_{k:k'-1}$. The probability of detection $\overline{p}_{a_{k:k'}}^{D,s,i}$ is given by \eqref{eqn: expected pd} using $\overline{x}_{k'|k',a_{k:k'}}^{i,h_{k:k'}^{i}}$ and $P_{k'|k',a_{k:k'}}^{i,h_{k:k'}^{i}}$. Note that \eqref{eqn: observing a measurement density multi-sensor} is calculated for the Bernoulli components that are present at time step $k$ and also for the Bernoulli components born between time step $k$ and $k'$.

Let $r_{k'|k',a_{k:k'}}^{i,h_{k:k'}^{i}}$, $P_{k'|k',a_{k:k'}}^{i,h_{k:k'}^{i}}$ be the probability of existence and covariance matrix of the $i$-the Bernoulli at time step $k'$ given the sequence of actions, $a_{k:k'}$ and observations $h_{k:k'}^{i}$. Then, using \eqref{eqn: simplified cost notation}, the MSGOSPA bound at time step $k'$ for actions $a_{k:k'}$ and sequence of detections/misdetections  $h_{k:k'}^i$ for the $i$-th Bernoulli is
\begin{multline} \label{eqn: expected MSGOSPA cost multi-sensor}
    \mathrm{C}^i\left(a_{k:k'},h_{k:k'}^{i}\right)  =\\
    \begin{cases}
    \frac{c^{2}}{2}r_{k'|k',a_{k:k'}}^{i,h_{k:k'}^{i}} & r_{k'|k',a_{k:k'}}^{h_{k:k'}}\leq\Gamma_{d}^{*}\\
    \frac{c^{2}}{2}\left(1-r_{k'|k',a_{k:k'}}^{i,h_{k:k'}^{i}}\right)\\
    +r_{k'|k',a_{k:k'}}^{i,h_{k:k'}^{i}}\min\left(\mathrm{tr}\left(P_{k'|k',a_{k:k'}}^{i,h_{k:k'}^{i}}\right),c^{2}\right) & r_{k'|k',a_{k:k'}}^{h_{k:k'}}>\Gamma_{d}^{*}.
    \end{cases}
\end{multline}

Then, the optimal policy in \eqref{eqn: J not nested} under the assumptions in Section \ref{sec: assumptions of the sensor management algorthm} can be written as the nested optimisations
\begin{align} \label{eqn: J nested new notation multi-sensor}
    \hat{J}_{\mu_{k:K}^{*}\left(\cdot\right)} &=  \nonumber \\
    & \underset{a_{k}}{\min}\sum_{i=1}^{n_{k|k}}\sum_{h_{k}^{i}\in\{0,1\}^{S}}\bigg[p(h_{k}^{i}|a_{k})\mathrm{C}^i\left(a_{k},h_{k}^{i}\right) \nonumber \\
     & \quad +\lambda\underset{a_{k+1}}{\min}\sum_{i=1}^{n_{k|k}}\sum_{h_{k+1}^{i}\in\{0,1\}^{S}}\bigg[p(h_{k+1}^{i}|a_{k:k+1},h_{k}^{i}) \nonumber \\
      & \quad\quad\quad\quad\quad\quad\quad\quad\quad\quad \times\mathrm{C}^i\left(a_{k:k+1},h_{k:k+1}^{i}\right)\ \nonumber \\
     & \quad+... \nonumber \\
     & +\lambda^{K-k+1}\underset{a_{K}}{\min}\sum_{i=1}^{n_{k|k}}\sum_{h_{K}^{i}\in\{0,1\}^{S}}p(h^i_{K}|a_{k:K-1},h_{k:K-1}^{i}) \nonumber\\
     &\quad\quad\quad\quad\quad\quad\quad\quad \times\mathrm{C}^i\left(a_{k:K},h_{k:K}^{i}\right)\bigg]...\bigg].
\end{align}

\section{Monte Carlo Tree Search Implementation}\label{sec: MCTS}
This section explains the MCTS implementation \cite{Browne2012} of the non-myopic multi-sensor multi-Bernoulli sensor management algorithm presented in the previous section. The MCTS implementation for Bernoulli filtering is described in \cite{Jones2024TAES}. Section \ref{sec: tree structure} explains the hypothesis tree and the  reduction we consider. Section \ref{sec: node information} indicates the information that is kept in each node of the tree. Section \ref{sec:MCTS} explains the MCTS implementation of the algorithm. Finally, Section \ref{sec:applicability} discusses the applicability of the designed algorithm.

\subsection{Tree Hypothesis Reduction}\label{sec: tree structure}
The MCTS algorithm aims to approximately solve \eqref{eqn: J nested new notation multi-sensor} by selecting the action with the lowest cost. To do so, MCTS creates a tree of actions and detections/misdetections for each sensor. In this case, the number of child nodes per parent is $2^S\times|\mathbb{A}|^S$, where $|\mathbb{A}|$ is the number of actions per sensor. This represents the number of actions per sensor multiplied by the number of possible detection/misdetection hypotheses per each sensor.

In this paper, we propose to simplify the tree structure by combining all detection and misdetection hypotheses for the same Bernoulli into a single hypothesis. This results in the tree only representing actions, not observation states, as done in \cite{Salvagnini2015}. This results in a tree where the number of child nodes (at the following time step) per parent is $|\mathbb{A}|^S$, instead of $2^S\times|\mathbb{A}|^S$. This also implies that there is only one type of edge in the tree, representing actions, while the original tree has two types of edges, representing actions and detections/misdetections. In addition, for computational efficiency and to be able to use the MSGOSPA bound in Lemma 1, we approximate each Bernoulli single-target density, which is a Gaussian mixture, as a Gaussian. We proceed to explain how to compute these Gaussian Bernoulli densities and the associated costs in the reduced tree.

Since the reduced tree only considers actions, the $i$-th Bernoulli is characterised by the parameters $r_{k'|k',a_{k:k'}}^{i}$, $\overline{x}_{k'|k',a_{k:k'}}^{i}$ and $P_{k'|k',a_{k:k'}}^{i}$. That is, these parameters do not depend on the sequence of observations $h_{k:k'}^{i}$ (as in the implementation with the entire tree).  When we perform an update with a predicted Bernoulli of this form, following Section \ref{sec: multi sensor update}, there is a Bernoulli density for each
$h_{k'}^{i}$ (sequence of detection misdetections) of the form
\begin{multline}
    f_{k'|k',a_{k'}}^{i,h_{k'}^{i}}\left(X\right)  =\\
    \begin{cases}
        r_{k'|k',a_{k:k'}}^{i,h_{k'}^{i}}\mathcal{N}\left(x;\overline{x}_{k'|k',a_{k:k'}}^{i,h_{k'}^{i}},P_{k'|k',a_{k:k'}}^{i,h_{k'}^{i}}\right) & X=\left\{ x\right\} \\
        1-r_{k'|k',a_{k:k'}}^{i,h_{k'}^{i}} & X=\emptyset \\
        0 & \mathrm{otherwise}
    \end{cases}
\end{multline}
where $r_{k'|k',a_{k:k'}}^{i,h_{k'}^{i}}$, $\overline{x}_{k'|k',a_{k:k'}}^{i,h_{k'}^{i}}$
and $P_{k'|k',a_{k:k'}}^{i,h_{k'}^{i}}$ are computed via \eqref{eqn: x0}-\eqref{eqn: r1}. 

Then, for each target, there is a mixture of these
Bernoulli densities, weighted by the corresponding probability of
$h_{k'}^{i}$ such that its density is
\begin{multline}\label{eq:Bernoulli_mixture}
    f_{k'|k',a_{k'}}^{i}\left(X\right)  = 
     \sum_{h_{k'}^{i}\in\{0,1\}^{S}}p(h_{k'}^{i}|a_{k:k'-1})f_{k'|k',a_{k'}}^{i,h_{k'}^{i}}\left(X\right)
\end{multline}
where $p(h_{k'}^{i}|a_{k:k'-1})$ is computed using \eqref{eqn: observing a measurement density multi-sensor} with probability of existence $r_{k'|k',a_{k:k'}}^{i}$. 

A mixture of Bernoulli densities is another Bernoulli density. In this case, its single-target density is a Gaussian mixture. To 
speed up the algorithm and be able to use the MSGOSPA bound in Lemma 1, we approximate this Gaussian mixture as a single Gaussian via moment matching. This is a standard procedure in target tracking algorithms, originating from the probabilistic data association filter \cite{Bar-Shalom09} and the joint probabilistic data association filter \cite{Bar-Shalom75}. The resulting Bernoulli density that is stored at each node of the tree is characterised by a mean $\overline{x}_{k'|k',a_{k:k'}}^{i}$, a covariance matrix $P_{k'|k',a_{k:k'}}^{i}$ and a probability of existence $r_{k'|k',a_{k:k'}}^{i}$. The formulas to calculate these parameters are provided in Appendix \ref{appendix_B} for completeness. 

The cost at time step $k'$ for the $i$-th Bernoulli for each $h^i_{k'}$ is denoted by $\mathrm{C}^i\left(a_{k:k'},h_{k'}^{i}\right)$, which is given by substituting $r_{k'|k',a_{k:k'}}^{i,h_{k'}^{i}}$ and $P_{k'|k',a_{k:k'}}^{i,h_{k'}^{i}}$ into $\eqref{eqn: expected MSGOSPA cost multi-sensor}$. Then, from \eqref{eqn: J nested new notation multi-sensor}, the cost at time step $k'$ (without considering the time-discounting factor $\lambda$) for the node representing action $a_{k:k'}$ is the weighted sum of the costs for each $h^i_{k}$ such that
\begin{multline}\label{eqn: cost at timestep k'}
   \mathrm{C}_{k'}(a_{k:k'})= \\
    \sum_{i=1}^{n_{k'|k}}\sum_{h_{k'}^{i}\in\{0,1\}^{S}}p(h_{k'}^{i}|a_{k:k'-1},h_{k:k'-1}^{i})\mathrm{C}^i\left(a_{k:k'},h_{k'}^{i}\right)
\end{multline}
where $\mathrm{C}\left(a_{k:k'},h_{k:k'}^{i}\right)$ is given by \eqref{eqn: expected MSGOSPA cost multi-sensor}. With this simplification, the number of children nodes per parent is $|\mathbb{A}|^{S}$.

\begin{figure}[h]
    \centering
    \includegraphics[width=1\linewidth]{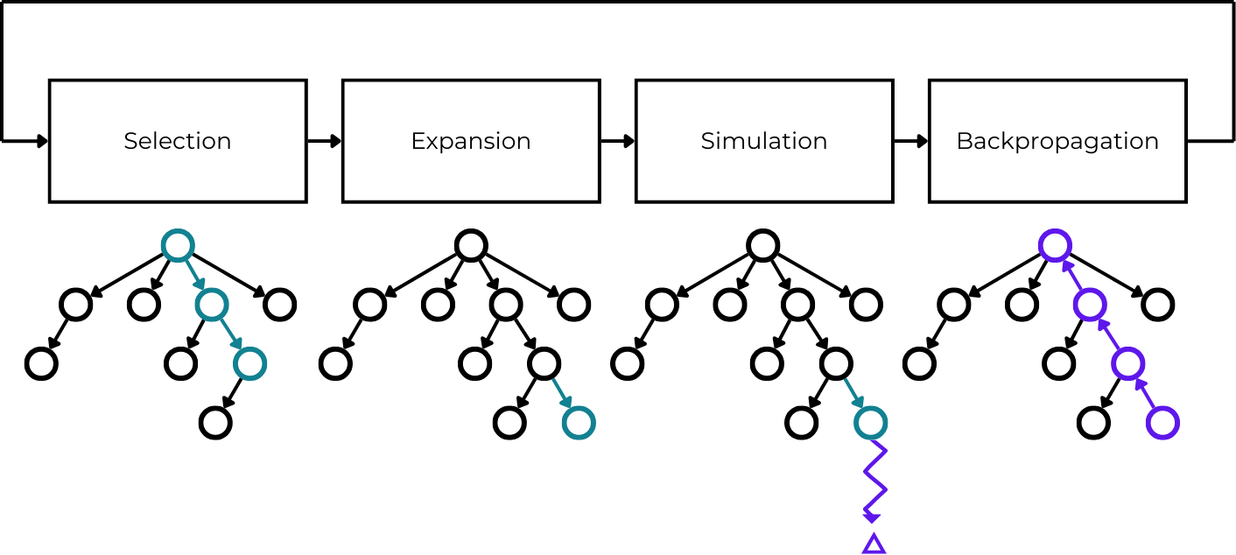}
    \caption{The four stages of the MCTS algorithm. Where $\Delta$ is the cost returned from the simulation phase. Figure adapted from \cite{BostroemRost2021} \cite{Browne2012}.}
    \label{fig: MCTS}
\end{figure}

\subsection{Node Information} \label{sec: node information}
Each node of the tree, representing a sequence of actions $a_{k:k'}$ up to time step $k'$ contains the following information:
\begin{itemize}
    \item Visit count $n$.
    \item Expected initial cost of visiting node $\mathrm{C}_{k'}$, given by \eqref{eqn: cost at timestep k'} (not updated by simulation outcomes).
    \item Expected cost of visiting node over all time steps $\mathrm{\overline{C}}_{k'}$, see \eqref{eqn: update node cost} (updated based on simulation outcomes).
    \item Sensors current locations.
    \item Parent node.
    \item Set of child nodes $J$.
    \item Set of available actions $\mathbb{A}_{k'}$.
    \item Target means $\overline{x}_{k'|k', a_{k:k'}}^i$, covariance matrices $P_{k'|k', a_{k:k'}}^i$ and probabilities of existence $r_{k'|k', a_{k:k'}}^i$ for $i\in\{1,...,n_{k'|k'}\}$, see \eqref{eqn: merged mean}-\eqref{eqn: merged prob existence}.
    \item Expected probabilities of detection for all targets and sensors $\overline{p}^{D,s,i}_{a_{k:k'}}$, see \eqref{eqn: expected pd}.
    \item Node depth in global tree $k'$.
\end{itemize}

It should be noted that at each time step $k'$, we add the Bernoulli components corresponding with the birth process.

\subsection{MCTS Algorithm} \label{sec:MCTS}
The MCTS algorithm is a selective and iterative way of exploring a search space. In sensor management, each row of depth in the tree can be understood as a time step into the future, where the root node represents the current time step.

There are four phases to the MCTS algorithm which are outlined below and illustrated in Figure \ref{fig: MCTS}. One MCTS iteration consists of these four stages and the budget refers to how many iterations are completed when building the tree.

\subsubsection{Selection}
This phase always begins at the root node. The selection criteria (commonly the upper confidence bound for trees (UCT) \eqref{eqn: UCT}) \cite{Browne2012} is used to continuously select children nodes, until a node is reached that does not have a full set of children. That is, we choose the node
\begin{equation}\label{eqn: UCT}
    \underset{j \in J}{\text{arg max}} \bigg\{\overline{\mathrm{C}}_{k', j} +  \epsilon \sqrt{\frac{\ln n}{n_j}} \bigg\}
\end{equation}
where $\epsilon$ is the trade-off parameter between exploration and exploitation, $n_j$ represents the visit count of the child node and $n$ the visit count of the current node, $J$ is the set of children of the current node, and $\overline{\mathrm{C}}_{k', j}$ is the expected cost of visiting the child node, as further explained in the backpropagation section.

Once a node has been selected that meets this criteria, we move to the expansion phase.

\subsubsection{Expansion}
Here, a new child node $j$ is added to the node selected in the previous stage of the algorithm. To add this node, we randomly select a previously untried action, predict the set of target densities (and add in the densities from the birth model), generate the ideal measurement set, calculate the cost of selecting this action and then update the set of target densities based on the consideration of both the detection and misdetection hypotheses \eqref{eqn: x0}-\eqref{eqn: r1} for each target. We can then use the results of these calculations to intialise a new node within the tree structure. The budget of the tree is the maximum number of nodes that can be added to the tree. Each time a node is added to the tree via the expansion phase, the budget of the tree is decremented by one.

\subsubsection{Simulation}
Starting from the node that was added during the expansion phase, a rollout is begun, in which actions are selected in-line with the rollout policy, which is often random. The rollout phase continues to move down the depth of the tree until the terminating criteria is met. This is often a user-specified maximum search depth. Once the rollout phase has been completed, a total, time-discounted cost of selecting this path of actions (beginning at the root node) is obtained, denoted as $\Delta$. During the simulation phase, no nodes are added to the tree and therefore the budget is not decremented by the simulation phase. We calculate $\Delta$ as
\begin{equation}\label{eqn: simulation delta}
    \Delta = \sum_{\kappa=k'+1}^{K}\lambda^{\kappa-k}\mathrm{C}_{\kappa}\left(a_{k:\kappa}\right)
\end{equation}
where $K$ is the length of the action path (which is also the total depth of the tree as it includes the simulation) and $\kappa$ indicates the depth of the corresponding node (and cost) in that tree.

\subsubsection{Back-propagation}
The total, time discounted cost of the path of used nodes $\Delta$ is then incorporated into the cost of each of the nodes (already in the tree) that are on the path. The costs are incorporated as a mean average \eqref{eqn: update node cost}, using the visit count $n$ of each node, yielding

\begin{equation}\label{eqn: update node cost}
    \overline{\mathrm{C}}_{k', new} = \frac{(\overline{\mathrm{C}}_{k', old} \cdot n) +\Delta}{n + 1}
\end{equation}
where $\overline{\mathrm{C}}_{k', old}$ is the cost prior to the backpropagation phase, $\overline{\mathrm{C}}_{k', new}$ refers to the updated cost associated with this action/node, $n$ is the visit count of the node, and $\Delta$ is the cost calculated during the simulation phase. After the cost is updated, the visit count is incremented by one.

Finally, the pseudo code of the GOSPA driven, non-myopic, multi-sensor management algorithm for multi-Bernoulli filtering is provided in Algorithm \ref{alg: ms, mb GOSPA driven sensor management}. 
\begin{algorithm}[H]
\caption{Non-myopic, multi-agent, multi-Bernoulli GOSPA driven sensor management}\label{alg: ms, mb GOSPA driven sensor management}

\begin{algorithmic}[1]

    \State Initialise tree with root node statistics 

    \While{budget is remaining}
        \State Select child node using UCT \eqref{eqn: UCT} $\gets$ \textsc{Selection}.
        \State Expand tree by adding node and calculating  $\overline{x}_{k'|k', a_{k:k'}}^{i}, P_{k'|k', a_{k:k'}}^{i}$ and $r_{k'|k', a_{k:k'}}^{i}$, using \eqref{eqn: x0}-\eqref{eqn: r1}, with the mixture merging  \eqref{eqn: merged prob existence}-\eqref{eqn: merged cov}, and        
        also considering new born targets \eqref{eqn: birth prob}-\eqref{eqn: birth density}.
        \State Compute the cost $\mathrm{C}_{k'}(a_{k:k'})$ using \eqref{eqn: cost at timestep k'} and add it to the node $\gets$ \textsc{Expansion}.
        \State Perform a simulation until terminating criteria is met, in accordance with rollout policy, obtaining the cost of the rollout $\Delta$ using \eqref{eqn: simulation delta} $\gets$ \textsc{Simulation}.
        \State Update $\overline{\mathrm{C}}_{k'}$ using \eqref{eqn: update node cost} and update the number of visit counts $n$ for all nodes on the action path $\gets$ \textsc{Back-propagation}.
    \EndWhile
         
        \State Select the action $a_k^s$ which has the lowest cost $\overline{\mathrm{C}}_{k'}$ \eqref{eqn: update node cost}.
    
\end{algorithmic}
\end{algorithm}

\subsection{Discussion on applicability}\label{sec:applicability}
The developed multi-target multi-sensor management algorithm can be used for any problem that can be modelled with the multi-target dynamic and measurement models in Section \ref{sec: Dynamic and measurement model}. That is, it can be used for any multi-sensor management problem with point targets. While the algorithm has been written for linear-Gaussian models, the extension to non-linear model is straightforward using linearisation techniques, which are widely used in Gaussian filtering \cite{Sarkka_book23}. For instance, the algorithm can be used to manage a set of sensors (e.g. drones) with a limited field of view that can move in the surveillance area. Another possible use of the algorithm is in cognitive radar \cite{Huang24}, where different waveform parameters can lead to different probabilities of detection or single-measurement models. It is also possible to use this algorithm to control a telescope for space situation awareness \cite{Oakes2022}.

\section{Simulations}\label{sec: simulations}
In this section, we provide simulation results with two sensors in a scenario in which obstacles block the movement of the sensors. All units in this section are in the international system but have been omitted for brevity. 

In Section \ref{sec: models}, the target motion model and sensor model are described, in Section \ref{sec: tree search considerations} the implementation details of the MCTS are provided and in Section \ref{sec: results}, we provide the simulation results.

\subsection{Models}\label{sec: models}
Each of the individual targets motion is governed by a nearly constant velocity model \cite[Chap. 6]{BarShalom2004}. The state vector is defined as $x = [p_x, v_x, p_y, v_y]^T$ in which $p$ denotes the position, $v$ denotes the velocity and subscript $x$ and $y$ denote the directional element of each of the vector quantities w.r.t. a two dimensional Cartesian grid. The transition and process noise covariance matrices are
\begin{equation}\label{eqn: transition and process noise matrices}
    F = I_2 \otimes 
\begin{pmatrix}
    1 & \tau \\
    0 & 1
\end{pmatrix}, \quad
    Q = q I_2 \otimes 
\begin{pmatrix}
    \tau^3 / 3 & \tau^2 / 2 \\
    \tau^2 / 2 & \tau
\end{pmatrix}
\end{equation}
where the sampling time $\tau = 1$ and $q = 0.8$. We consider a Bernoulli birth density that is parameterised by the mean $\overline{x}_B = [0, 0.1, 0, 0.1]^T$, covariance $P_B = \text{diag}([6, 6, 6, 6]^T)$ and probability of birth $r_k^b = 0.03$. That is, we know within reasonable accuracy the locations where targets may be born. The probability of survival is $p^S = 0.99$ and each simulation is $200$ time steps in length.

\begin{figure}[h]
    \centering
    \includegraphics[width=0.8\linewidth]{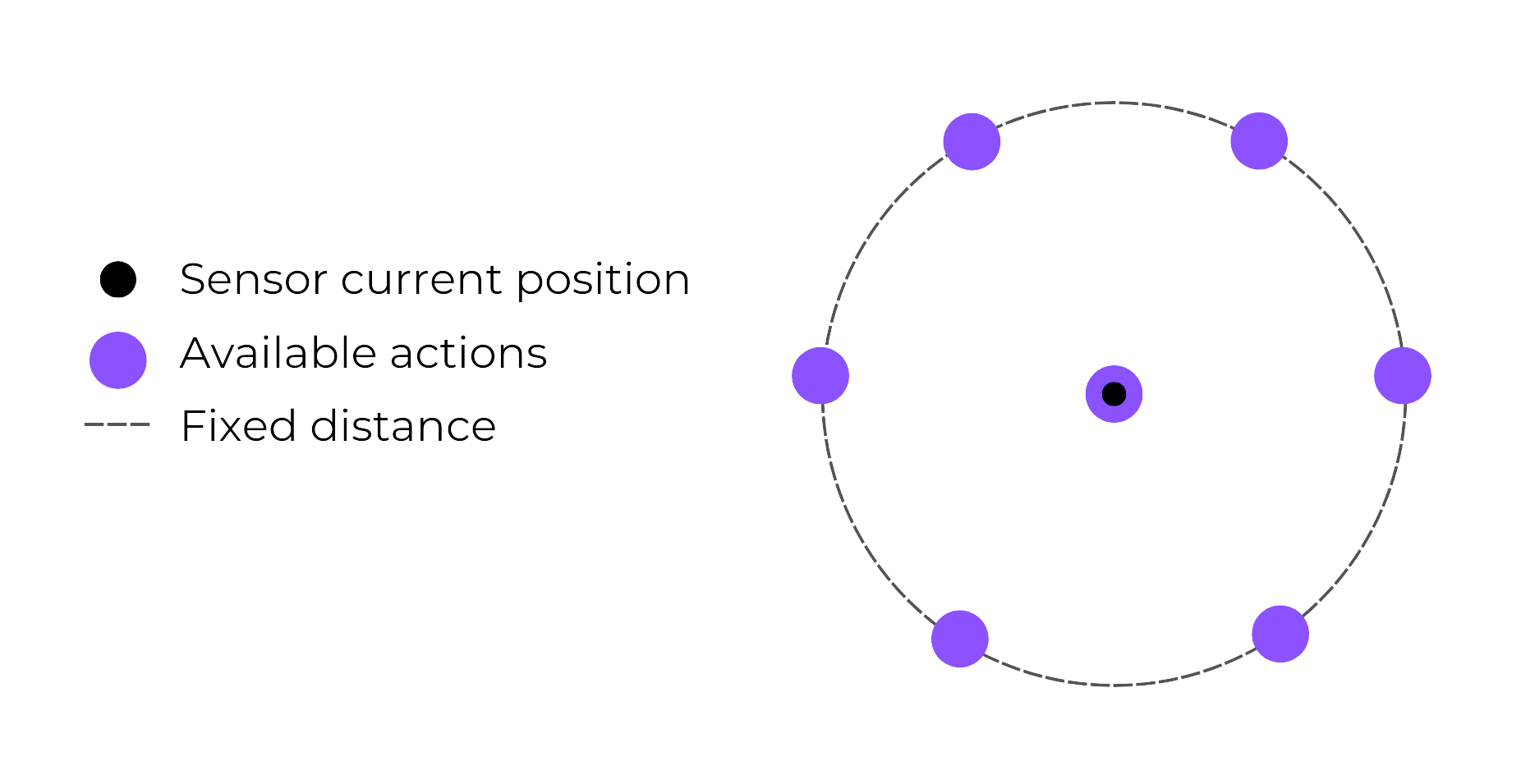}
    \caption{Sensor movement model, showing seven possible actions for an individual sensor. Six actions cause the sensor to move and one action cause the sensor to hold its position.}
    \label{fig: sensor actions}
\end{figure}

We consider the case of $S=2$ sensors. Each sensor has seven discrete actions to select from at each time step. The actions determine where the sensor will move at the next time step. Six of these actions are positioned around the circumference of a circle of constant radius $15$, centered at the senors current position; the seventh action is for the sensor to remain in its current position \ref{fig: sensor actions}. These actions can be understood as changes in direction with a constant velocity, with the added option of not moving.

The sensors collect imperfect measurements, including noise on detections of the ground truth and also clutter. The probability of detection for a target at a distance $\delta$ from sensor $s$ is
\begin{equation}
    p^{D,s}_{a_k} = p^D  e^{-\frac{1}{2}  (\frac{\delta}{r})^2}
\end{equation}
where $p^D = 0.999$ and $r=40$. That is, the probability of detection decays exponentially with distance. The expected value of the probability of detection is approximated by its value at the predicted mean. The observation matrix $H$ and observation noise matrix $R$ are 
\begin{equation} \label{eqn: observation and observation noise matrices}
H^s_{a_k} = \begin{bmatrix}
    1 & 0 & 0 & 0\\
    0 & 0 & 1 & 0\\
\end{bmatrix} ; 
R^s_{a_k} =  \begin{bmatrix}
    2 & 0\\
    0 & 2\\
\end{bmatrix}
\end{equation}
with a bias term $b^s_{a_k}=0$. Clutter is generated by a PPP and uniformly distributed in the sensors FOV, which has a radius of $r=40$. The clutter rate is $\overline{\lambda}^C_{a_k}= 0.1$.

The surveillance region is a square of size $500\times500$. The movements of the sensor platforms are restricted by the obstacle as actions that breach their boundary are not available for selection.

In the simulations, the sensors are initialised such that the obstacle blocks their direct path to the birth location - meaning in all scenarios, the obstacle has to be overcome. This way, in every simulation that is ran, the sensor has to navigate around the obstacle in order to meet its operational objective of tracking the unknown and time varying number of targets. 

An obstacle-free scenario has been omitted from the results but it can be noted that there is no major benefit for adopting computationally intense non-myopic approaches for these simple scenarios.

\subsection{Tree Search Considerations} \label{sec: tree search considerations}
To speed up the algorithm, if the sensors are closer than a proximity threshold $\Psi$, then we build an MCTS for both sensors to optimise over both sensor actions jointly. If they are farther away, we build an MCTS for each sensor that is optimised independently for each sensor.

It can be mentioned here that if the proximity threshold $\Psi$ is too small, in this scenario, the sensors continuously move between inside and outside of the threshold, which is not helpful for planning. The value of $\Psi = 3r$ has been selected in these simulations and prevents this from happening, whilst also allowing for the computational benefits of not jointly optimising at every time step.

For the scenario in which we jointly optimise the two sensors, we build a tree in which each parent node can have a maximum of $7^2 = 49$ children, meaning that each child node represents an action pair, one relating to each sensor. Conversely, in the scenario where the sensors are independently optimised, each parent node has a maximum of $7$ children.

\subsection{Results}\label{sec: results}
We evaluate several variations of the MCTS implementation of the GOSPA driven (GD) sensor management algorithm\footnote{The Python code of the proposed algorithm will be available at \mbox{\url{https://github.com/sggjon5}} after publication.}. We also use the same MCTS tree structure for a set of information theoretic sensor management algorithms, driven by the KLD \cite{Kreucher07}. This algorithm aims to maximise the sum of the KLDs between the predicted densities and the posterior Bernoulli densities for each target \cite[Appendix C]{Jones2024TAES}. 

For the MCTS implementations, we vary the computational budgets (number of nodes allowed to be added to the tree via expansion) and lookahead for the MCTS (MCTS1-4). The parameters tested in the simulations are provided in Table \ref{tab: algorithm parameters}, including the budgets for joint and individual optimisation, see Section \ref{sec: tree search considerations}. The reason for differing budgets for the two types of optimisation are due to the search spaces being larger in the joint optimisation case, and therefore a proportionally larger budget is provided. The decay factor ($\lambda=0.9$) and distance threshold of joint optimisation ($\Psi = 3r$) where $r$ is the radius of the sensors FOV, have been held constant for these simulations. We also include two myopic algorithms (Myopic - GD, Myopic - KLD) in which the parameters of the MCTS are set such that there is no rollout and the tree cannot grow beyond the first layer.
\begin{table}[h]
\centering
\caption{Algorithm parameters for GD and KLD.}
\label{tab: algorithm parameters}
\begin{tabular}{@{}lccc@{}}  
\toprule
\textbf{Algorithm} & \textbf{Budget - Joint} & \textbf{Budget - Individual} & \textbf{Lookahead} \\ \midrule
Myopic & 49 & 7 & 1 \\
MCTS1 & 49 & 7 & 5 \\
MCTS2 & 49 & 7 & 10 \\
MCTS3 & 200 & 40 & 5 \\
MCTS4 & 200 & 40 & 10 \\ \bottomrule
\end{tabular}
\end{table}

In this implementation, we use loopy belief propagation \cite{Williams2014} to compute the MB projection and we merge Bernoulli components after each update if their Mahalanobis distance is smaller than $1$ \cite[Eqs. (25-31)]{Fontana2023}. We merge similar Bernoulli components because of computational efficiency. This does however eventually cause the birth density to surpass the threshold of being declared as an existing target. Once this threshold of existence is surpassed, the errors (if within the maximum localisation error distance) become localisation errors. 

\begin{figure}
    \centering
    \includegraphics[width=\linewidth]{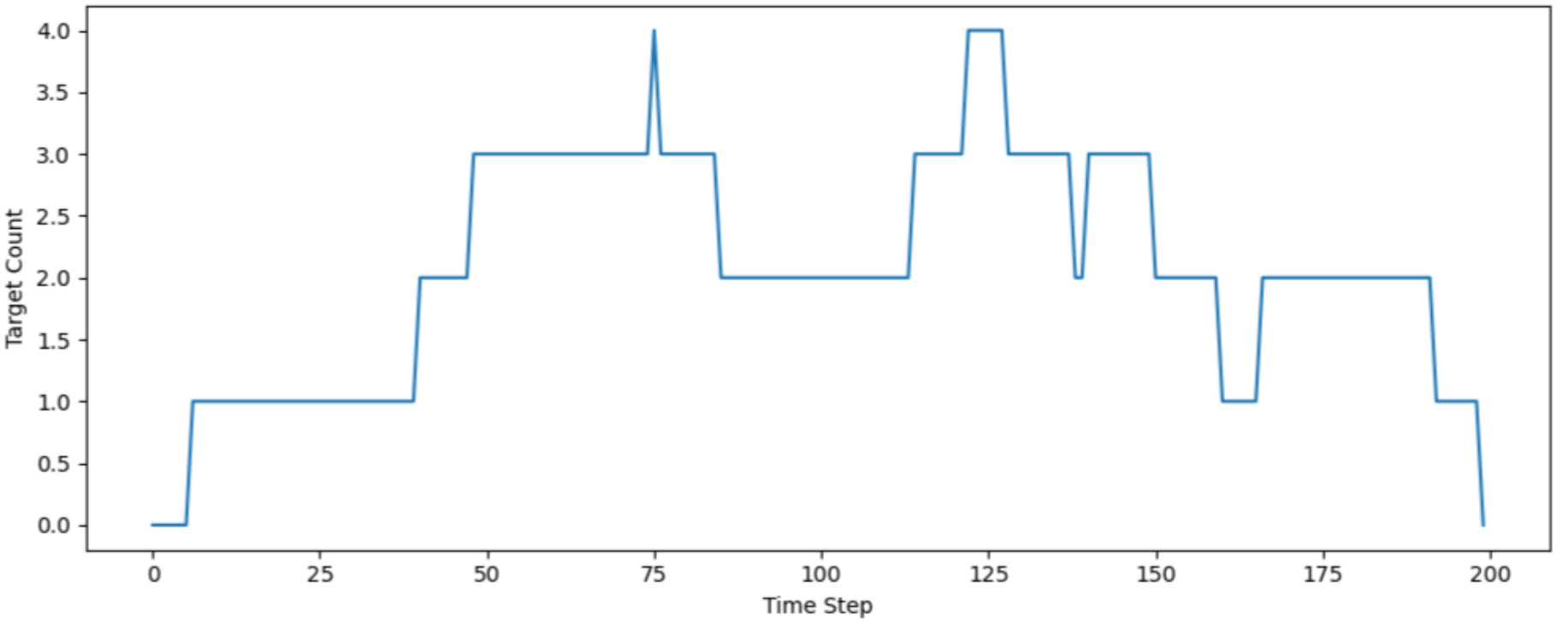}
    \caption{Number of targets alive at each time step in the simulation. Starting with zero targets and a maximum number of four at time steps 75 and 125. The total number of targets in the simulation is 8.}
    \label{fig: target count}
\end{figure}

The root mean square GOSPA (RMS-GOSPA) errors were calculated over a simulation of 200 time steps and 50 Monte Carlo runs. The GOSPA parameter $c = 2r$, where $r = 40$ is the radius of the sensor's field of view (FOV), and the parameter $p = 2$. The planning horizon is set to $N$, meaning the action space has a maximum of $7^N$ options to search through when optimized individually, and ${7^{S N}}$ options when jointly optimized, where $N_s$ is the number of sensors being jointly optimized.  The number of targets alive at each time step of the simulation can be seen in Figure \ref{fig: target count}, the most targets alive at any one time is four.

In this simulation, to ensure that the sensor management algorithms have to overcome the obstacle, when each run is initialised, the sensors direct path to observe the birth location (the only known Bernoulli component in the MB filter at conception), is blocked. This way, the algorithms must navigate around an obstacle to begin viewing the birth location. Due to this initialisation criteria, myopic algorithms quickly evidence their shortcomings, as they are unable to navigate around the obstacles and therefore get stuck. It should be noted here that all of the algorithms often miss the first target as they are initialised some distance from the birth location for the reasons described above.  

\begin{figure}[h]
    \centering
    \includegraphics[width=0.9\linewidth]{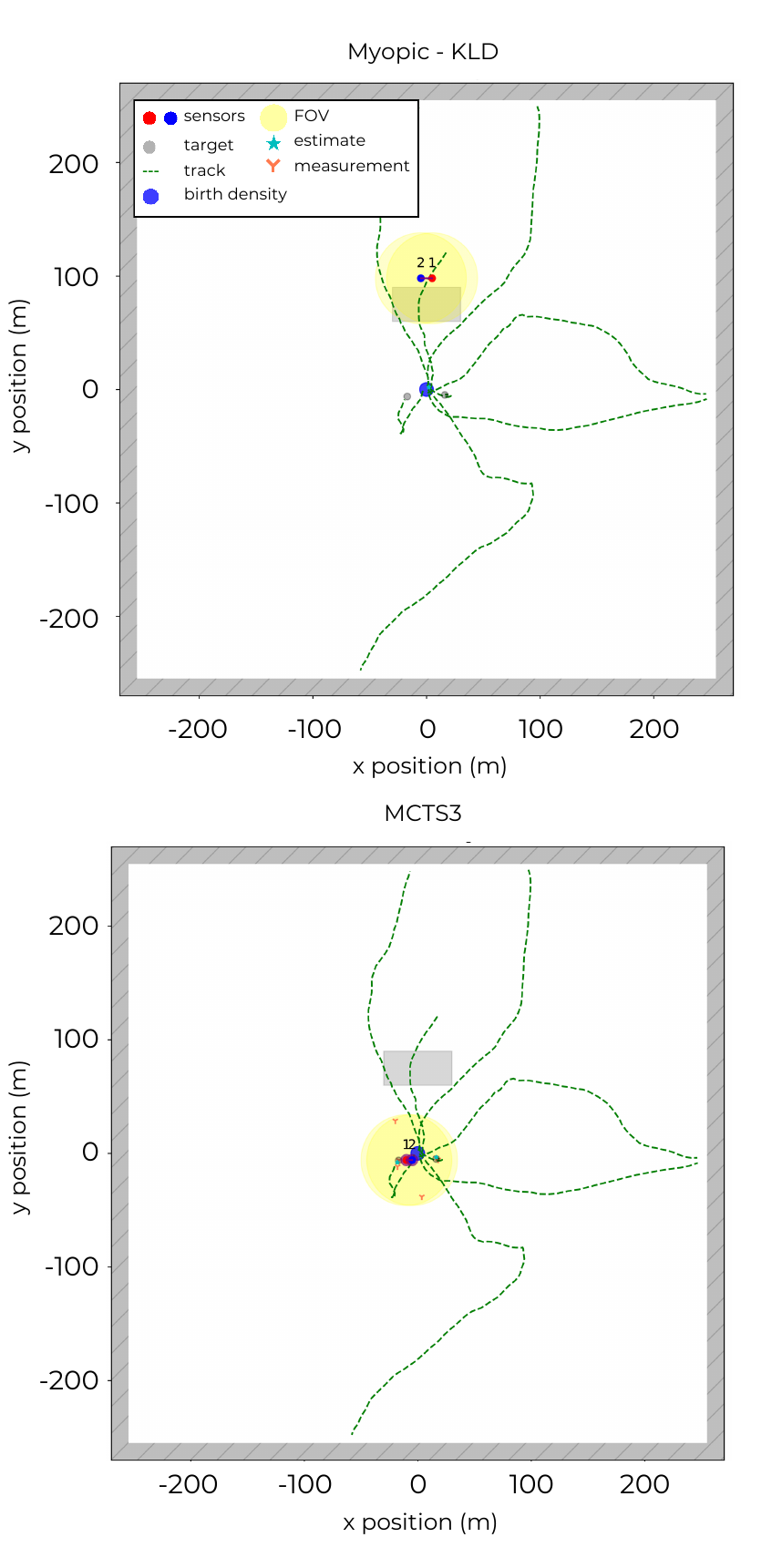}
    \caption{Frame 152 in one Monte Carlo run, number of targets alive: 2. Top - myopic KLD scenario snapshot. Both sensors are stuck behind the obstacle as they do not have the ability to plan further ahead. 
    Bottom - MCTS3 scenario snapshot. Sensors have navigated around the obstacle and are maintaining track of the two targets alive at this time step. Purple outline on each sensor indicates they are planning with a non-myopic policy, purple line between the two sensors indicates they are being jointly optimised at this time step. Birth density is centred at the origin.}
    \label{fig:scenario screenshots}
\end{figure}

Figure \ref{fig:scenario screenshots} shows the outputs from the myopic KLD planning algorithm, and the MCTS3 - GD algorithm at time step $152$ in a Monte Carlo run. In the myopic (KLD and GD) algorithms, both sensors get stuck behind the obstacle.  In the non-myopic case (MCTS1-4), the sensors do not get stuck behind the obstacle. In the MCTS3 - GD results, both sensors are near the birth area, maintaining track of the two targets currently alive. The purple outline on the sensors indicates that they are non-myopically planning and the purple line between the sensors indicates that the sensors are jointly optimised.


\begin{figure}[h]
    \centering
    \includegraphics[width=1\linewidth]{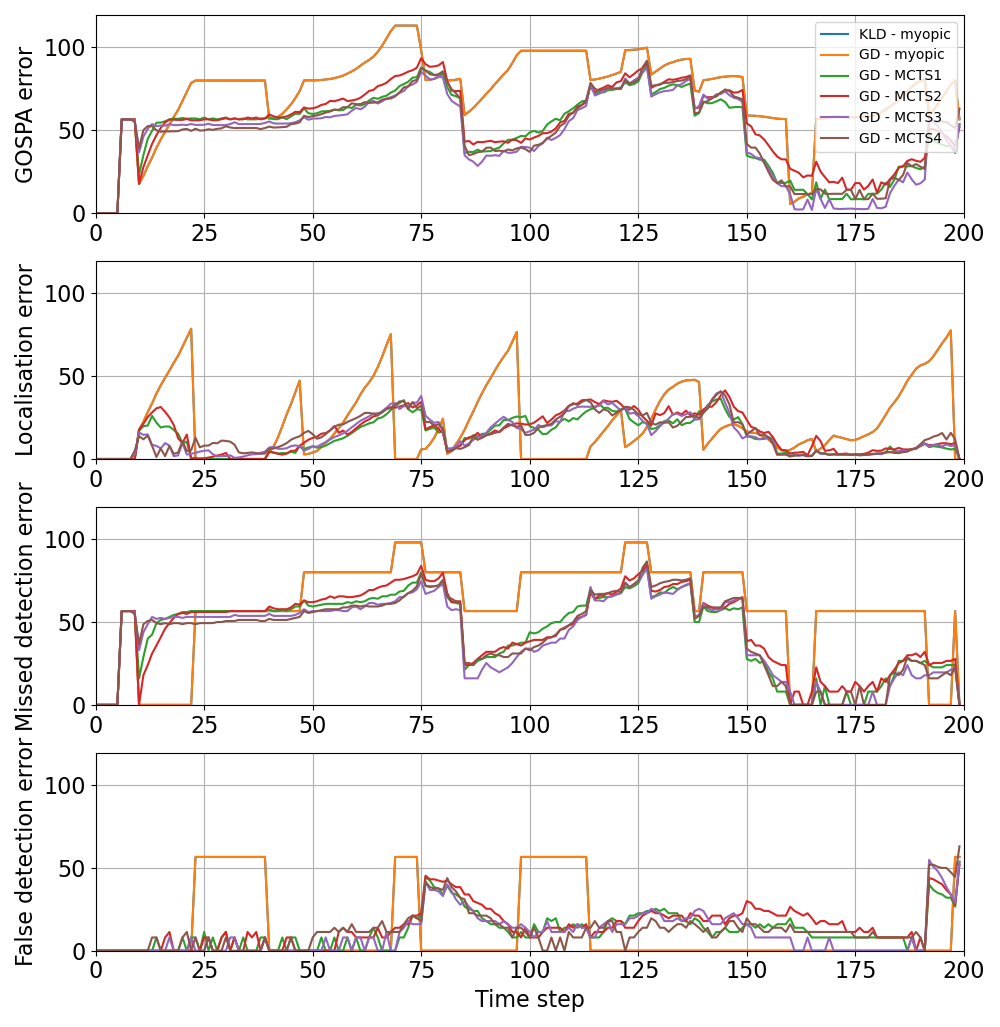} %
    \caption{GOSPA error breakdown for the obstacles scenario. The top plot shows the overall GOSPA error across each MC run for each time step. The second, third and fourth plots show the localisation, missed and false errors that contribute to the overall GOSPA error. Result set for  $\overline{\lambda}^C_{a_k} = 1$.}
    \label{fig: GOSPA error - obstacles}
\end{figure}

Figure \ref{fig: GOSPA error - obstacles} shows the RMS-GOSPA error at each time step. For clarity in the figure presentation, the errors for the non-myopic KLD algorithm are not shown here. As expected, non-myopic algorithms improve upon the myopic alternatives as they can navigate around the obstacle. In particular, myopic algorithms get stuck behind the obstacle and the errors are identical. MCTS3 - GD is the best performing algorithm in this scenario, outperforming the other non-myopic algorithms, even those with a longer lookahead. This is likely due to the longer lookahead allowing for increased uncertainty in the simulation phase and therefore the predictions that inform the sensor actions contain higher uncertainty; finding this balance of the appropriate lookahead we found to be dependent on the scenario. By this, we mean that the relationship between the size of the obstacles being navigated around and the distance covered by the lookahead length need to be commensurate. For example, if an obstacles is ten units square, and the sensor platform moves in steps of two, we would suggest a planning horizon of at least five to allow the algorithms to propose paths that navigate around the obstacle. In Figure \ref{fig:scenario screenshots}, there are two targets currently alive, and a third has just died. This is reflected in Figure \ref{fig: GOSPA error - obstacles} by a sudden drop in the missed target error as at this point the target that died was not being tracked. 

It should also be noted here that the large contribution of localisation and missed target errors in the myopic planning algorithm are caused by the merging of similar Bernoulli components at the birth location within the MB filter \cite{Fontana2023}. The localisation error for the myopic algorithms can be understood to increase as the targets move away from the birth area, with a sharp drop off when the targets either die, or they are re-categorised as missed targets once they reach the maximum localisation error $c$.

Table \ref{tab: RMSGOSPA error - obstacles} shows the average RMS-GOSPA for each algorithm across all time steps for three differing clutter rates $\overline{\lambda}^C_{a_k} \in \{0.1, 1, 2\}$. From this we can see that the myopic algorithms have consistently higher error throughout the simulation as they are unable to navigate around the obstacle within the surveillance area and therefore unable to meet their operational objective. In addition, performance is quite consistent across these different clutter rates. It is also worth noting that the non-myopic GOSPA driven approach outperforms the KLD approach in all but one setting (MCTS2 with $\overline{\lambda}^C_{a_k}= 2$). We believe this may be due to the fact that it is not explicitly clear what maximising for information gain means in a multi-target tracking context, but minimising localisation, missed and false target errors (GOSPA error), is.
\begin{table}[h]
\centering
\caption{RMS-GOSPA error across all time steps for each algorithm over 50 MC runs with varying clutter rates.}
\label{tab: RMSGOSPA error - obstacles}
\begin{tabular}{@{}lccc@{}}  
\toprule
\textbf{Algorithm} & $\overline{\lambda}^C_{a_k}= 0.1$ & $\overline{\lambda}^C_{a_k}= 1$ & $\overline{\lambda}^C_{a_k}= 2$ \\ \midrule
Myopic - GD   & 72.80 & 72.80 & 72.80 \\
MCTS1 - GD    & 49.05 & 50.46 & 51.78 \\
MCTS2 - GD    & 54.59 & 54.00 & 56.18 \\
MCTS3 - GD    & 47.17 & 47.88 & 50.53 \\
MCTS4 - GD    & 48.24 & 49.96 & 51.75 \\ 
\midrule
Myopic - KLD & 72.80 & 72.80 & 72.80 \\
MCTS1 - KLD    & 51.94 & 53.16 & 53.16 \\
MCTS2 - KLD    & 54.74 & 54.71 & 54.71 \\
MCTS3 - KLD    & 51.38 & 51.51 & 51.51 \\
MCTS4 - KLD    & 51.76 & 52.36 & 52.36 \\

\bottomrule
\end{tabular}
\end{table}
\begin{table}[h]
\centering
\caption{Wall clock run time of each algorithm.}
\label{tab: algorithm runtimes}
\begin{tabular}{@{}lcc@{}}  
\toprule
\textbf{Algorithm} & \textbf{Driven by} & \textbf{Runtime per time step [s]} \\ 
\midrule
Myopic & GOSPA & 0.02 \\
MCTS1  & GOSPA & 0.23 \\
MCTS2  & GOSPA & 0.62 \\
MCTS3  & GOSPA & 0.85 \\
MCTS4  & GOSPA & 2.57 \\ 
\midrule
Myopic & KLD & 0.02 \\
MCTS1  & KLD & 0.23 \\
MCTS2  & KLD & 0.61 \\
MCTS3  & KLD & 0.85 \\
MCTS4  & KLD & 2.56 \\
\bottomrule
\end{tabular}
\end{table}

The simulations were ran on an Intel Core i7-14700KF 3.4GHz CPU and the wall clock run time of each algorithm ($\overline{\lambda}^C_{a_k} = 2$) is given in Table \ref{tab: algorithm runtimes}. This computational time remains basically unaltered for the three considered clutter rates. As we would expect, the algorithms with a larger lookahead and a larger computational budget (more nodes in the tree) take longer to process each time step, whilst the myopic approaches are the least computationally demanding.

\section{Conclusions}\label{sec: conclusion}
We have proposed a sensor management algorithm that is transparent, multi-sensor and multi-Bernoulli, driven by the GOSPA metric. Utilising GOSPA as the cost function in sensor management, provides a desirable level of interpretability  and transparency within the decision making. This is because the GOSPA-derived cost function specifically optimises for the quantities of interest in a multi-target tracking scenario.

The proposed approach has general applicability to non-myopic multi-Bernoulli sensor management algorithms. We have demonstrated the benefits of non-myopic planning when operating in a surveillance area that contains obstacles.

A line of future research is to develop GOSPA-based non-myopic multi-target multi-sensor management algorithm with fewer approximations. These should have better performance, though with a higher computational complexity. Another line of future research is the development of reinforcement learning algorithms for multi-target multi-sensor management, for instance, using a reward based on the GOSPA metric.
\bibliographystyle{elsarticle-num} 
\bibliography{References.bib}

\clearpage

{\LARGE Supplementary material: GOSPA-Driven Non-Myopic Multi-Sensor Management with Multi-Bernoulli Filtering}{\LARGE\par}

\appendices
\section{}\label{appendix_A}
In this appendix, we prove Lemma 1 by deriving the MSGOSPA error in a single time step, which is obtained by setting $K=k$ in \eqref{eqn: J not nested} (myopic planning). Using \eqref{eqn: predicted density of measurement sequence} and \eqref{eqn: MSGOSPA for given MB posterior} and integrating over all measurements, the MSGOSPA error is 
\begin{multline}
    E =\int\bigg[\int d^{2}\left(X_{k},\hat{X}_{k}\left(a_{k},Z_{k}\right)\right)\\
    \times f_{k|k}\left(X_{k}|Z_{k};a_{k}\right)\delta X_{k}\bigg] \prod_{s=1}^{S}f_{k|k-1}^{m,s}\left(Z_{k}^{s}\right)\delta Z_{k}^{1:S}.
\end{multline}

Using \eqref{eqn: cost inequality bernoulli components}, we obtain the upper bound $B$ of $E$ $(E \leq B)$ with
\begin{align}
    B  &= \int\dots\int  \left[\sum_{i=1}^{n_{k|k}}\mathrm{C}\left(r_{k|k,a_{k}}^{i,|Z_{k}^{i}|},P_{k|k,a_{k}}^{i,|Z_{k}^{i}|}\right)\right]\nonumber \\
    & \times\prod_{s=1}^{S}\left[f_{k|k-1}^{m,p,s}\left(Z_{k}^{s,0}\right)\prod_{i=1}^{n_{k|k-1}}f_{k|k-1}^{m,i,s}\left(Z_{k}^{s,i}\right)\right] \nonumber \\
    & \quad\quad\quad\quad\quad\quad\quad\quad    \left[\prod_{s=1}^{S}\prod_{i=0}^{n_{k|k-1}}\delta Z_{k}^{s,i}\right].
\end{align}
Then, since the cost does not depend on $Z_k^{s,0}$ we can integrate the PPP densities out (each integral has value one). Then, the integral of the sum can be written as the sum of the integrals. This yields
\begin{align}\label{eqn: B part two}
    B & =\int\dots\int\left[\sum_{i=1}^{n_{k|k}}\mathrm{C}\left(r_{k|k,a_{k}}^{i,|Z_{k}^{i}|},P_{k|k,a_{k}}^{i,|Z_{k}^{i}|}\right)\right]\nonumber \\
     & \quad\quad\times\prod_{s=1}^{S}\left[\prod_{i=1}^{n_{k|k-1}}f_{k|k-1}^{m,i,s}\left(Z_{k}^{s,i}\right)\right]\left[\prod_{s=1}^{S}\prod_{i=1}^{n_{k|k-1}}\delta Z_{k}^{s,i}\right]\nonumber \\
     & =\sum_{i=1}^{n_{k|k}}\int\dots\int\mathrm{C}\left(r_{k|k,a_{k}}^{i,|Z_{k}^{i}|},P_{k|k,a_{k}}^{i,|Z_{k}^{i}|}\right) \nonumber \\
     & \quad\quad\quad\quad \prod_{s=1}^{S}f_{k|k-1}^{m,i,s}\left(Z_{k}^{s,i}\right)\left[\prod_{s=1}^{S}\delta Z_{k}^{s,i}\right].
\end{align}

Now we plug \eqref{eqn: predicted density of the measurement} into \eqref{eqn: B part two} and carry out the set integrals \cite{Mahler_book14}. This results in the MSGOSPA error bound in Lemma 1 that is used in the proposed sensor management algorithm
\begin{align}\label{eqn: B upper bound}
    & B =\sum_{i=1}^{n_{k|k}}\sum_{|Z_{k}^{1,i}|=0}^{1}...\sum_{|Z_{k}^{S,i}|=0}^{1}\mathrm{C}\left(r_{k|k,a_{k}}^{i,|Z_{k}^{i}|},P_{k|k,a_{k}}^{i,|Z_{k}^{i}|}\right)\nonumber \\
     & \quad\times\prod_{s=1}^{S}\bigg[\left(1-|Z_{k}^{s,i}|\right)\left(1-r_{k|k-1}^{i}\overline{p}_{a_{k}}^{D,s,i}\right) \nonumber \\
     & \quad\quad\quad\quad\quad\quad\quad +|Z_{k}^{s,i}|r_{k|k-1}^{i}\overline{p}_{a_{k}}^{D,s,i}\bigg].
\end{align}

\section{}\label{appendix_B}
In this appendix, we provide the details of the Gaussian mixture merging used to approximate the single-target density in the Bernoulli density \eqref{eq:Bernoulli_mixture}, which is a Gaussian mixture, as a Gaussian. 

First of all, the probability of existence of the Bernoulli \eqref{eq:Bernoulli_mixture} is the weighted sum of the probability of existence of each Bernoulli
\begin{align}
    r_{k'|k',a_{k:k'}}^{i} & =\sum_{h_{k'}^{i}\in\{0,1\}^{S}}p(h_{k'}^{i}|a_{k:k'-1})r_{k'|k',a_{k:k'}}^{i,h_{k'}^{i}}. \label{eqn: merged prob existence}
\end{align}
Then, the merged mean and covariance matrix are given by the standard moment matching formula for Gaussian mixtures, where the weight of each Gaussian mixture component is proportional to $p(h_{k'}^{i}|a_{k:k'-1})r_{k'|k',a_{k:k'}}^{i,h_{k'}^{i}}$, resulting in \cite{BarShalom2004}

\begin{align}
    \overline{x}_{k'|k',a_{k:k'}}^{i} & = \frac{1}{r_{k'|k',a_{k:k'}}^{i}}\sum_{h_{k'}^{i}\in\{0,1\}^{S}}p(h_{k'}^{i}|a_{k:k'-1}) r_{k'|k',a_{k:k'}}^{i,h_{k'}^{i}} \nonumber \\
    & \quad\quad\quad\quad\quad\quad\quad\quad \times \overline{x}_{k'|k',a_{k:k'}}^{i,h_{k'}^{i}} \label{eqn: merged mean} \\
    P_{k'|k',a_{k:k'}}^{i} & =\frac{1}{r_{k'|k',a_{k:k'}}^{i}}\sum_{h_{k'}^{i}\in\{0,1\}^{S}}p(h_{k'}^{i}|a_{k:k'-1}) r_{k'|k',a_{k:k'}}^{i,h_{k'}^{i}} \nonumber \\ & \times  \bigg[P_{k'|k',a_{k:k'}}^{i,h_{k'}^{i}}+\left(\overline{x}_{k'|k',a_{k:k'}}^{i,h_{k'}^{i}}-\overline{x}_{k'|k',a_{k:k'}}^{i}\right) \nonumber\\
     & \quad\quad\quad\quad\times\left(\overline{x}_{k'|k',a_{k:k'}}^{i,h_{k'}^{i}}-\overline{x}_{k'|k',a_{k:k'}}^{i}\right)^{T}\bigg]. \label{eqn: merged cov}
\end{align}

\end{document}